\documentclass[conference]{IEEEtran}

\IEEEoverridecommandlockouts                              % This command is only needed if 
\usepackage{subfig}
\usepackage[lined,ruled]{algorithm2e}		
\makeatletter

\newcommand{\Rmnum}[1]{\expandafter\@slowromancap\romannumeral #1@}
\makeatother																	

\usepackage{epsfig} % for postscript graphics files
\usepackage{amsmath} % assumes amsmath package installed
		\usepackage{caption}

\title{\LARGE \bf
Model-Driven Data Collection for Biological Systems
}

\author{\IEEEauthorblockN{Xiao Lin}
\IEEEauthorblockA{Department of Computer Science\\
University of South Carolina\\
Columbia, South Carolina 29205\\
Email: lin65@email.sc.edu}
\and
\IEEEauthorblockN{Gabriel Terejanu}
\IEEEauthorblockA{Department of Computer Science\\
University of South Carolina\\
Columbia, South Carolina 29205\\
Email: terejanu@cec.sc.edu}
}

\begin{document}

\maketitle
\thispagestyle{empty}
\pagestyle{empty}

%%%%%%%%%%%%%%%%%%%%%%%%%%%%%%%%%%%%%%%%%%%%%%%%%%%%%%%%%%%%%%%%%%%%%%%%%%%%%%%%
\begin{abstract}

For biological experiments aiming at calibrating models with unknown parameters, a good experimental design is crucial, especially for those subject to various constraints, such as financial limitations, time consumption and physical practicability. In this paper, we discuss a sequential experimental design based on information theory for parameter estimation and apply it to two biological systems. Two specific issues are addressed in the proposed applications, namely the determination of the optimal sampling time and the optimal choice of observable. The optimal design, either sampling time or observable, is achieved by an information-theoretic sensitivity analysis. It is shown that this is equivalent with maximizing the mutual information and contrasted with non-adaptive designs, this information theoretic strategy provides the fastest reduction of uncertainty.

\end{abstract}

%%%%%%%%%%%%%%%%%%%%%%%%%%%%%%%%%%%%%%%%%%%%%%%%%%%%%%%%%%%%%%%%%%%%%%%%%%%%%%%%
\section{INTRODUCTION}

The understanding of underlying dynamics in biological systems through mathematical models requires a good parameter estimation from experimental data. However, experimental data collection in biological systems is a time consuming and costly operation due to its high reliance on human resources and specialized equipment. As a result, strategies based on design of experiments are needed to judiciously perform data collections that meet various constraints of biological experiments~\cite{Kreutz:2009vq}.

In this paper, we adopt an information-theoretic sequential experimental design which reduces the uncertainty in parameters by selecting designs that maximize the expected information gain. This is equivalent with finding the experimental design that can provide the highest statistical dependence between the model parameters and observables~\cite{Terejanu:2011wx}. The study shows that by monitoring the reduction in uncertainty after each measurement update, one can design a stopping criteria for the experimental process which gives a minimal set of experiments to efficiently learn the model parameters.

The sequential model-driven data collection approach is an iteration between optimal design determination, measurement collection and parameter estimation. This is naturally posed in the Bayesian framework. After each optimal design is determined, Bayesian update is applied to obtain the posterior distribution of model parameters given the newly acquired experimental data. We then repeat the cycle by using the current posterior distribution to determine the next optimal design.

The basis for our data collection strategy is Shannon entropy~\cite{Shannon:2001wk}, which quantifies the uncertainty of a random variable. Lindley~\cite{Lindley:1956wj} proposes that the amount of information provided by an experiment can be quantified as the difference between the entropy of the prior and the entropy of the posterior. To anticipate the information gain before performing the experiment, Lindley proposes to average the information gain over all possible experimental outcomes. This expected information gain is interpreted as mutual information~\cite{Terejanu:2011wx,Paninski2005}. Thus the optimal experimental design is fount by maximizing the mutual information. 

Sebastiani and Wynn~\cite{Sebastini2000} show that when the conditional entropy of the observable given the parameters is not functionally dependent on the design then the optimal design can be found by just maximizing the entropy of the observable. This maximum entropy strategy is a particular case of mutual information maximization. Even though estimating the entropy is easier than estimating the mutual information, we show that for the problems studied in this paper, this independence assumption does not hold. Therefore we adopt the general mutual information maximization strategy for our data collection process.

Estimating mutual information from samples is challenging. Some commonly used estimators include histogram based estimator, kernel density estimator, and $k$-nearest neighbor estimator (kNN). In their survey, Walters-Williams and Li~\cite{WaltersWilliams:2009gh} show that parametric estimation usually outperform non-parametric estimation when data is drawn from a known family of distributions. However, this is not the case in most practical problems. Khan et al.~\cite{Khan:2007up} compare different estimators that quantify the dependence between random variables, and show that the kNN estimator of mutual information captures better the nonlinear dependence than other commonly used estimators. In our paper, we adopt the kNN estimator of mutual information proposed by Kraskov et al.~\cite{Kraskov:2004gr}, which is based on kNN estimator of entropy proposed by Kozachenko and Leonenko~\cite{Kozachenko:1987ts}. 

After each design stage a Bayesian update needs to be performed to obtain the posterior distribution of the parameters. Given that the biological systems in this paper are described by nonlinear ordinary differential equations, Markov Chain Monte Carlo (MCMC) and sequential Monte Carlo methods are commonly used to solve the Bayesian update problem~\cite{Vanlier:2013ux}. Since we need to sequentially solve for the posterior, we adopt the ensemble Kalman filter (EnKF) to obtain samples that are distributed according to the posterior distribution of model parameters. The EnKF propagates a relatively small ensemble of samples through the system nonlinearities and during the measurement update it moves the ensembles such as their mean and covariance approximate the first two moments of the posterior.
 
The rational behind our sequential data collection strategy is described in Section~\ref{sec:exp_design}. Bayesian update is detailed in Section~\ref{sec:bayesian_update}, and in Section~\ref{sec:applications} we apply this sequential experimental design to two biological models: predator-prey and cell-signaling pathway. Finally, the conclusions and future work are given in Section~\ref{sec:conclusions}.

%----------------------------

\section{INFORMATION-THEORETIC EXPERIMENTAL DESIGN}
\label{sec:exp_design}

In our work, the goal of experiments is to estimate unknown parameters of biological systems. Each observation contributes to the reduction in parametric uncertainty. Design of experiments determine experimental conditions which can provide the observations that carry the most information leading to the least uncertainty in the posterior parameters.

The biological models used in this paper are generally represented by the following ordinary differential equations:
\begin{align}
  &\dot{\textbf{x}}(t) = \textbf{f}(\textbf{x}(t),\boldsymbol{\theta},\boldsymbol{\xi}) \label{eq_process}\\
  &\textbf{d} = \textbf{H}\textbf{x}(t) + \boldsymbol{\epsilon}_{meas} \label{eq_meas}
\end{align}
where $\textbf{x}(t)$ is the state of the system $\boldsymbol{\xi}\in\boldsymbol{\Xi}$ is the design variables associated with the experimental scenario, and $\boldsymbol{\Xi}$ is the design space. The uncertain model parameters are denoted by $\boldsymbol{\theta}\in\boldsymbol{\Theta}$, and the random vector $\boldsymbol{\epsilon}_{meas}$ is the observation noise due to measurement imprecision. The measurement noise is assumed to be normally distributed with zero mean and covariance matrix $\textbf{R}$. The observable $\textbf{d}$ is assumed here to be a linear combination of system states. The measurement data is obtained at a series of observation times, and the parameters $\boldsymbol{\theta}$ are estimated from the observations via Bayesian update.

In the followings we present an information-theoretic experimental design, to determine the experimental condition $\boldsymbol{\xi}^*$ that provides the highest reduction in parametric uncertainty.

\subsection{Criterion to find the optimal design}

According to Lindley~\cite{Lindley:1956wj}, the amount of information provided by an experiment $\boldsymbol{\xi}$, with prior knowledge $p(\boldsymbol{\theta)}$, when the observation is $\textbf{d}$, is
\begin{equation}
  \begin{aligned}
    U(\textbf{d},\boldsymbol{\xi}) = &-\int_{\boldsymbol{\Theta}}p(\boldsymbol{\theta})\log{p(\boldsymbol{\theta})}d\boldsymbol{\theta} -  \\
    &(-\int_{\boldsymbol{\Theta}}p(\boldsymbol{\theta}|\textbf{d})\log{p(\boldsymbol{\theta}|\textbf{d})}d\boldsymbol{\theta}) \label{information gain}
  \end{aligned} 
\end{equation}
where the two terms on the right represent the entropy of prior and posterior distribution respectively. The utility function $U(\textbf{d},\boldsymbol{\xi})$ quantifies the information gained from the experimental data $\textbf{d}$. However, since we do not know the observation before performing the experiment, we compute the average amount of information provided by an experiment $\boldsymbol{\xi}$ by marginalizing over the all possible observables predicted by the model:
\begin{equation}
E_{d}[U(\textbf{d},\boldsymbol{\xi})] = \int_{\textbf{D}}U(\textbf{d},\boldsymbol{\xi})p(\textbf{d}|\boldsymbol{\xi})d\textbf{d} \label{expected information gain}
\end{equation}
Given the above expected information gain, the optimal experiment is obtained by solving the following optimization problem:
\begin{equation}
\boldsymbol{\xi}^{*} = arg \max\limits_{\boldsymbol{\xi}\in\boldsymbol{\Xi}} E_{d}[U(\textbf{d},\boldsymbol{\xi})]
\end{equation}
We can expand the expected information gain by substituting Eq.~\eqref{information gain} in Eq.~\eqref{expected information gain}:
\begin{align}
E_{d}[U(\textbf{d},\boldsymbol{\xi})]&= \int_{\textbf{D}}\int_{\boldsymbol{\Theta}}p(\boldsymbol{\theta},\textbf{d}|\boldsymbol{\xi})\log\frac{p(\boldsymbol{\theta},\textbf{d}|\boldsymbol{\xi})}{p(\textbf{d}|\boldsymbol{\xi})}d\boldsymbol{\theta}d\textbf{d} \nonumber  \\
&\quad - \int_{\textbf{D}}\int_{\boldsymbol{\Theta}}p(\boldsymbol{\theta},\textbf{d}|\boldsymbol{\xi})\log{p(\boldsymbol{\theta})}d\boldsymbol{\theta}d\textbf{d} \\
&= \int_{\textbf{D}}\int_{\boldsymbol{\Theta}}p(\boldsymbol{\theta},\textbf{d}|\boldsymbol{\xi})\log\frac{p(\boldsymbol{\theta},\textbf{d}|\boldsymbol{\xi})}{p(\textbf{d}|\boldsymbol{\xi})p(\boldsymbol{\theta})}d\boldsymbol{\theta}d\textbf{d} \label{expanded MI}\\
&= I(\boldsymbol{\theta};\textbf{d}|\boldsymbol{\xi})
\end{align}
Here, $I(\boldsymbol{\theta};\textbf{d}|\boldsymbol{\xi})$ is the mutual information between model parameters $\boldsymbol{\theta}$ and model predictions/observables $\textbf{d}$. Mutual information provides a measure of statistical dependence between the parameters and observables and it quantifies the reduction in uncertainty in the parameters $\boldsymbol{\theta}$ when knowing $\textbf{d}$. Therefore, our model-driven data collection is based on mutual information maximization.
\begin{equation}
\boldsymbol{\xi}^{*} = arg \max\limits_{\boldsymbol{\xi}\in\boldsymbol{\Xi}} I(\boldsymbol{\theta};\textbf{d}|\boldsymbol{\xi})
\end{equation}
This information theoretic experimental design determines the experimental scenario $\boldsymbol{\xi}^{*}$ where the expected observations have the highest impact on model parameters.

%-------------

\subsection{Relation with maximum entropy maximization}
Note that Eq.~\eqref{expanded MI} can also be expanded as follows:
\begin{align}
  E_{d}[U(\textbf{d},\boldsymbol{\xi})]&= \int_{\textbf{D}}\int_{\boldsymbol{\Theta}}p(\boldsymbol{\theta},\textbf{d}|\boldsymbol{\xi})\log\frac{p(\boldsymbol{\theta})p(\textbf{d}|\boldsymbol{\theta},\boldsymbol{\xi})}{p(\textbf{d}|\boldsymbol{\xi})p(\boldsymbol{\theta})}d\boldsymbol{\theta}d\textbf{d} \\
  &= \int_{\textbf{D}}\int_{\boldsymbol{\Theta}}p(\boldsymbol{\theta},\textbf{d}|\boldsymbol{\xi})\log{p(\textbf{d}|\boldsymbol{\theta},\boldsymbol{\xi})}d\boldsymbol{\theta}d\textbf{d} \\
  &- \int_{\textbf{D}}\int_{\boldsymbol{\Theta}}p(\boldsymbol{\theta},\textbf{d}|\boldsymbol{\xi})\log{p(\textbf{d}|\boldsymbol{\xi})}d\boldsymbol{\theta}d\textbf{d} \\
  &= -H_{\xi}(\textbf{d}|\boldsymbol{\theta}) + H_{\xi}(\textbf{d}) \label{entropy}
\end{align}
where the first term in Eq.~\eqref{entropy} is the conditional entropy of the observable given the parameters, and the second term is the entropy of the observable.

When $H_{\xi}(\textbf{d}|\boldsymbol{\theta})$ is independent on the design variable $\boldsymbol\xi$, then the optimal design can be found by maximizing the entropy of the observable, $H_{\xi}(\textbf{d})$. Note that
\begin{equation}
  p(\textbf{d}|\boldsymbol{\theta},\boldsymbol{\xi}) = \int_{\textbf{X}}p(\textbf{d}|\textbf{x}(t))p(\textbf{x}(t)|\boldsymbol{\theta},\boldsymbol{\xi})d\textbf{x}(t) ~.
\end{equation}
For systems described by Eqs.~\eqref{eq_process}-\eqref{eq_meas}, that have
certain initial conditions, namely $\textbf{x}(0)$ is perfectly known, $p(\textbf{x}(t)|\boldsymbol{\theta})$ is a Dirac delta function and $p(\textbf{d}|\boldsymbol{\theta})$ will be solely determined by the distribution of the measurement noise. Hence $H_{\xi}(\textbf{d}|\boldsymbol{\theta}) = 1/2\log|2\pi{e}\textbf{R}|$ which is independent of the design variable. However, in the presence of uncertain initial conditions, $H_{\xi}(\textbf{d}|\boldsymbol{\theta})$ is in general dependent on the design variable, which means maximum entropy strategy will not be applicable. This is especially accentuated in sequential inference. Even though we may start with a certain initial condition in the beginning, due to parametric uncertainty, subsequently we end up with posterior distributions for the state which become priors in future designs. Thus we adopt mutual information maximization in our design strategy.

%-------------

\subsection{Mutual information estimation}
In this paper, mutual information is estimated using the kNN estimator proposed by Kraskov et al.~\cite{Kraskov:2004gr} as follows:
\begin{align}
I(\boldsymbol{\theta};\textbf{d}|\boldsymbol{\xi}) \approx & -\frac{1}{N}\sum^{N}_{i=1}\left(\psi(n_{\theta}(i) + 1) + \psi(n_{d}(i) + 1)\right) \nonumber \\
&+ \psi(k) + \psi(N)
\end{align}
Here, $n_{d}(i)$ and $n_{\theta}(i)$ are the number of samples in the marginal space within the distance from the $i$th sample to its kNN in the joint space, and $\psi(k)$ is the digamma function which satisfy the recursion $\psi(k+1) = \psi(k) + 1/x$ and $\psi(1) = -C$, where $C \approx 0.5772156$ is the Euler-Masheroni constant. Note that a small value for $k$ will result in a small bias but a large variance and vice-versa. Also, the efficiency of the estimator decreases as the dimensionality of the joint space increases. 

%----------------------------

\section{BAYESIAN UPDATE}
\label{sec:bayesian_update}

After performing an experiment, the observation data is collected, and the posterior distribution is obtained using Bayes rule. In this paper both state $\textbf{x}$ and parameter $\boldsymbol{\theta}$ are updated together. We treat $\boldsymbol{\theta}$ also as a state variable by augmenting the following equation to the process model in Eq.\eqref{eq_process},
\begin{equation}
  \dot{\boldsymbol{\theta}} = \textbf{0} ~. 
\end{equation}
For simplicity we denote with $\textbf{x}_\theta$ the vector of both state and parameters, $\textbf{x}_\theta = [\textbf{x} ~ \boldsymbol{\theta}]^T$. In sequential inference, the posterior distribution obtained at the current stage will be the prior distribution in the next stage. Suppose $\textbf{D}_{n}=\left\{\textbf{d}_{1}, \textbf{d}_{2},..., \textbf{d}_{n}\right\}$ is a set of observations at $n$ sequential stages. We can derive the following recursive Bayesian update:
\begin{align}
p(\textbf{x}_\theta|\textbf{D}_{n})&= \frac{p(\textbf{D}_{n}|\textbf{x}_\theta)p(\textbf{x}_\theta)}{p(\textbf{D}_{n})}\\
&= \frac{p(\textbf{d}_{n},\textbf{D}_{n-1}|\textbf{x}_\theta)p(\textbf{x}_\theta)}{p(\textbf{d}_{n},\textbf{D}_{n-1})}\\
&= \frac{p(\textbf{d}_{n}|\textbf{D}_{n-1},\textbf{x}_\theta)p(\textbf{D}_{n-1}|\textbf{x}_\theta)p(\textbf{x}_\theta)}{p(\textbf{d}_{n}|\textbf{D}_{n-1})P(\textbf{D}_{n-1})}\\
&= \frac{p(\textbf{d}_{n}|\textbf{D}_{n-1},\textbf{x}_\theta)p(\textbf{x}_\theta|\textbf{D}_{n-1})}{p(\textbf{d}_{n}|\textbf{D}_{n-1})}
\end{align}
Given the joint posterior distribution of state and parameters, the posterior distribution of the parameters is obtained by marginalization.

In this study the EnKF is used to obtain samples from posterior distributions. EnKF was first introduced by Evensen~\cite{Evensen:2009ud} and has been widely used in various applications. It originates from Kalman filter and it has similarities also with the particle filter (PF). Both EnKF and PF use samples to describe probability distributions; the samples are called ensemble member in EnKF and particles in PF.

EnKF updates each ensemble member at time $k$ as follows:
\begin{align}
  &\boldsymbol{x}^{a}_{\theta,j}(k) = \boldsymbol{x}^{f}_{\theta,j}(k) + \boldsymbol{\Sigma}^{f}_{e}(k)\textbf{H}^{T}[\textbf{H}\boldsymbol{\Sigma}^{f}_{e}(k)\textbf{H}^{T} + \textbf{R}_{e}]^{-1} \times \nonumber \\
&\quad\quad\quad [\textbf{d}_{j}(k) - \textbf{H}\boldsymbol{\Sigma}^{f}_{e}(k)] \\
  &\boldsymbol{\Sigma}^{f}_{e}(k) = \overline{[\textbf{x}^{f}_\theta(k) - \overline{\textbf{x}^{f}_\theta}(k)][\textbf{x}^{f}_\theta(k) - \overline{\textbf{x}^{f}_\theta}(k)]^{T}}\\
  &\boldsymbol{\Sigma}^{a}_{e} = \overline{[\textbf{x}^{a}_\theta(k) - \overline{\textbf{x}^{a}_\theta}(k)][\textbf{x}^{a}_\theta(k) - \overline{\textbf{x}^{a}_\theta}(k)]^{T}}
\end{align}
where the perturbed measurements $\textbf{d}_{j}(k) = \textbf{d}(k) + \boldsymbol{\epsilon}_{j}$, for $j=1\ldots N$ and $N$ is the number of ensemble members, and $\textbf{R}_{e} = \overline{\boldsymbol{\epsilon}\boldsymbol{\epsilon}^{T}}$ is the samples covariance matrix of the measurement noise. The analysis ensemble member $\boldsymbol{x}^{a}_{\theta,j}(k)$ represents the $j$th sample from the current posterior distribution. The forecast ensemble member $\boldsymbol{x}^{f}_{\theta,j}(k)$ is obtained by propagating the corresponding ensemble, $\boldsymbol{x}^{a}_{\theta,j}(k-1)$, from the previous posterior distribution through the mathematical model. For more details on theoretical analysis and implementation of EnKF one can refer to Ref.\cite{Evensen:2009ud}.

\section{APPLICATION}
\label{sec:applications}

For biological experiments, the design scenario can be the choice of observation times, pattern of stimulations, and/or the choice of observables \cite{Vanlier:2013ux}. In this section, maximum mutual information strategy is applied to two models: predator-prey and STAT5 cell-signaling pathway. For the first model, we focus on the optimal observation times, and for the second model we focus on the observable selection. 

\subsection{Predator-Prey}

The Lotka-Volterra equation
\begin{align}
&\dot{x}_{1}(t) = -\theta_{1}x_{1}(t)x_{2}(t) + \theta_{3}x_{1}(t) \\
&\dot{x}_{2}(t) = \theta_{2}x_{1}(t)x_{2}(t) - \theta_{4}x_{2}(t) 
\end{align}
is widely used to describe the dynamics of biological systems in which a prey and its predator interact. The states $x_{1}$ and $x_{2}$ are population of a prey and its predator, $\dot{x}_{1}(t)$ and $\dot{x}_{2}(t)$ represent the growth rates of the two populations over time and the parameters $\theta_{1}$ and $\theta_{2}$ describe the interaction of the two species. Suppose $\theta_{3}$ and $\theta_{4}$ are known, then in order to estimate parameters $\theta_{1}$ and $\theta_{2}$, sequential experiments are performed. In each experiment, $x_{1}(t)$ and $x_{2}(t)$ are measured with errors, and posterior distribution $p(\boldsymbol{\theta}|\textbf{x}_\theta(t))$ can be obtained via Bayesian update. This posterior distribution will be taken as the prior information for the following experiment. However, different measurement times will lead to different amount of reduction in parametric uncertainty. In our study, we compare maximum mutual information design strategy proposed in previous sections with non-adaptive strategies. 

\begin{figure}[htp]
 \small
  \centering
	\includegraphics[width=0.5\textwidth]{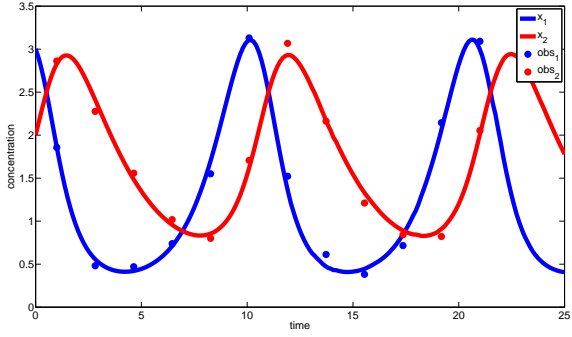}
\captionsetup{font={scriptsize}}
\caption{Curves show how population changes with time and discrete dots are the corresponding measurements simulated with true parameters and additive Gaussian noise.}
	\label{fig:pp} 
\end{figure}

For simulation, we use the following linear observation model with an additive Gaussian noise $\epsilon_{meas} \sim N(0,0.1^{2})$
\begin{equation}
d_{i} = x_{i}(t) + \epsilon_{meas},   \quad \quad \quad i = 1, 2
\end{equation}
We use normal priors $N(0.7, 0.1^{2})$ for $\theta_{1}$ and $N(0.4, 0.1^2)$ for $\theta_{2}$ and fix the other two parameters to $\theta_{3} = 1$, $\theta_{4} = 0.4$. The initial condition is given by $x_{1}(t_{0})=2$ and $x_{2}(t_{0})=3$. True parameters $[\theta_{1}, \theta_{2}] = [0.6, 0.3]$ are used to generate measurements, and simulated data is shown in Fig. \ref{fig:pp}. 

We divide the time span $[1, 21]$ into $4$ periods and in each period there are $3$ optional time points for measurement. The maximum mutual information strategy is employed to adaptively select a measurement time in each period. The reduction in the parametric uncertainty at the end of each period obtained using mutual information maximization is compared with non-adaptive strategies and a random strategy. The non-adaptive strategies always select the same measurement time in each period and the random strategy picks at random a measurement time. Here, the uncertainty of parameters is quantified using standard deviation for individual parameters and entropy for joint parametric distributions. We are looking for the most decrease in parametric uncertainty as more experiments are selected.

\begin{figure}[htp]
  \centering 
  \subfloat[\scriptsize{standard variance of $\theta_{1}$}]{ 
    \label{fig:subfig:a} %% label for first subfigure 
    \includegraphics[width=0.5\textwidth]{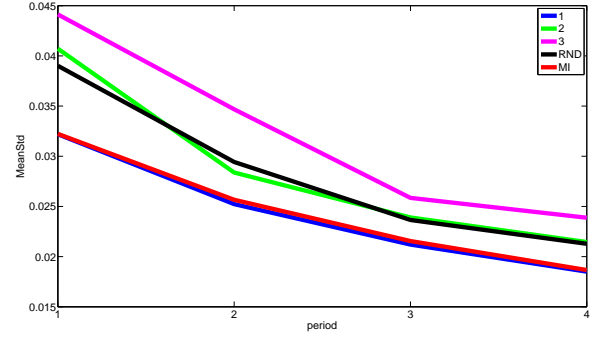}}
%		\caption{standard variance of $\theta_{1}$} 
		 \hspace{0.1in} 
	  \subfloat[\scriptsize{standard variance of $\theta_{2}$}]{ 
    \label{fig:subfig:b} %% label for first subfigure 
    \includegraphics[width=0.5\textwidth]{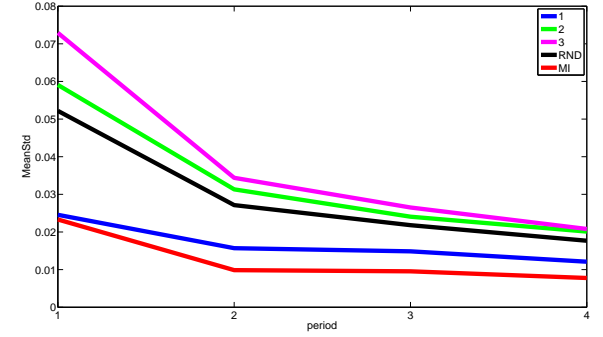}}
%		\caption{standard variance of $\theta_{2}$}
		 \hspace{0.1in} 
	  \subfloat[\scriptsize{joint entropy of $\theta_{1}$ and $\theta_{2}$}]{ 
    \label{fig:subfig:c} %% label for first subfigure 
    \includegraphics[width=0.5\textwidth]{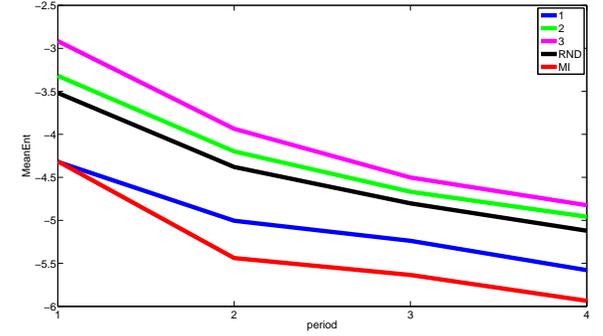}}
%		\caption{joint entropy of $\theta_{1}$ and $\theta_{2}$} 
		 \hspace{0.1in} 
  \subfloat[\scriptsize{root mean square error}]{ 
    \label{fig:subfig:d} %% label for second subfigure 
    \includegraphics[width=0.5\textwidth]{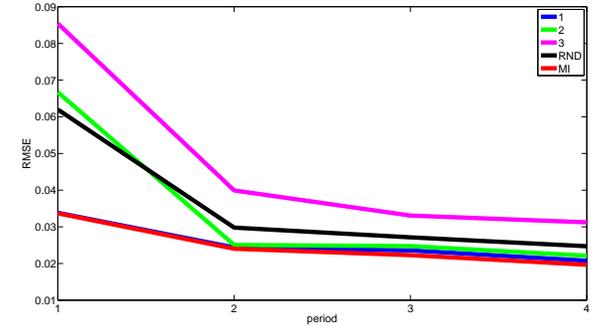}}
	%	\caption{ root mean square error}
		\captionsetup{font={scriptsize}}
  \caption{The performance of maximum mutual information strategy and non-adaptive strategies are compared. Figure (a) and Figure (b) show the standard variance of $\theta_{1}$ and $\theta_{2}$ after each experiment of different sequential designs. Figure (c) shows the joint entropy of $\theta_{1}$ and $\theta_{2}$ which quantifies the total uncertainty of both parameters. Figure (d) gives the root mean square error(RMSE) of parameter estimation} 
  \label{fig:results} %% label for entire figure 
\end{figure}

We use $1000$ ensemble members in the EnKF and to estimate the mutual information, and the performance measures are averaged over $100$ trial runs. Results are shown in Fig. \ref{fig:results}. We can see that strategy $1$, which always chooses the first measurement time in each period, and maximum mutual information strategy outperform the other strategies for reducing the uncertainty of $\theta_{1}$. However maximum mutual information provides the most reduction in the joint distribution of both parameters over all considered strategies. 

%-------------------

\subsection{STAT5 Cell-Signaling Pathway}

For the second example we use the STAT5 Cell-Signaling Pathway model presented by Swameye et al.~\cite{Swameye04022003}. For simplicity, we ignore the time delay and use a sustained stimulus. The model is given by the following set of ordinary differential equations:
\begin{align}
  &\dot{x}_{1}(t) = -\theta_{1}x_{1}(t)\\
  &\dot{x}_{2}(t) = -\theta_{2}x_{2}(t)^{2} + \theta_{1}x_{1}(t)\\
  &\dot{x}_{1}(t) = -\theta_{3}x_{3}(t) + \frac{1}{2}\theta_{2}x_{2}(t)^{2}\\
  &\dot{x}_{2}(t) = \theta_{3}x_{3}(t)
\end{align}
where $x_{1}(t)$ refers to the concentration of unphosphorylated STAT5, $x_{2}(t)$ is the activated STAT5, $x_{3}(t)$ is the dimeric STAT5, and $x_{4}(t)$ is the nucleus STAT5. The parameters $\theta_{1}$, $\theta_{2}$, $\theta_{3}$ are three reaction rates which need to be estimated from experimental data. 

Suppose that experimentally is more cost effective to measure either the total amount of activated STAT5 given by $y_{1} = x_{2} + 2x_{3}$ or STAT5 given by $y_{2} = x_{1} + x_{2} + 2x_{3}$. This means that every time step we have to make a choice between $y_{1}$ and $y_{2}$, and different observables may cause different amount of reduction in uncertainty. In order to estimate the reaction rates, sequential experiments are performed and mutual information strategy is employed to select which observable needs to be measured at each time step. 

\begin{figure}[htp]
  \centering 
  \subfloat[\scriptsize{STAT5 cell-signaling pathway model}]{ 
    \label{fig:subfig:c} %% label for first subfigure 
    \includegraphics[width=0.5\textwidth]{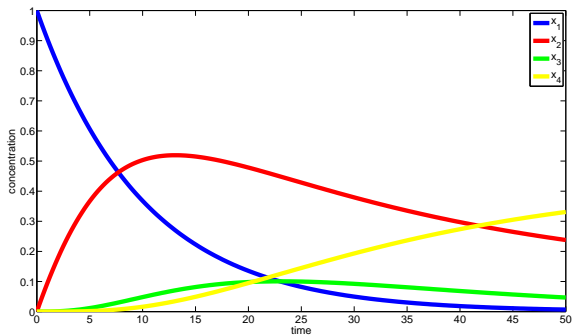}} 
  \hspace{0.1in} 
  \subfloat[\scriptsize{Observables}]{ 
    \label{fig:subfig:d} %% label for second subfigure 
    \includegraphics[width=0.5\textwidth]{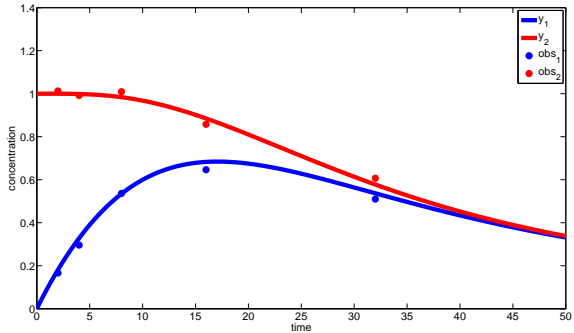}} 
		\captionsetup{font={scriptsize}}
  \caption{Figure(a) shows STAT5 cell-signaling pathway model. Figure(b) shows simulated time evolution of observables and corresponding measurements at time point $t = 2, 4, 8, 16, 32$.} 
  \label{SE} %% label for entire figure 
\end{figure}

In our simulation, the three parameters have the same Gaussian prior $N[0.5, 0.1]$, and the initial conditions are given by $x_{1}(0) = 1$, $x_{2}(0) = 0$, $x_{3}(0) = 0$, $x_{4}(0) = 0$. The true parameters $\theta_{1} = 0.1$, $\theta_{2} = 0.1$, $\theta_{3} = 0.1$ and the additive Gaussian noise $\epsilon_{meas} \sim N(0,0.1^{2})$ are used to simulate the system and generate measurements as shown in Fig.\ref{SE}.

Here we apply the maximum mutual information strategy to choose the optimal observable and compare it with non-adaptive strategies: such as always choosing the same observable and randomly choosing one observable at each time step. As in the previous example the uncertainty of parameters is quantified by standard deviation when referring to the marginal distributions and entropy for joint distributions. 

\begin{figure}[htp]
  \centering 
  \subfloat[\scriptsize{standard variance of $\theta_{1}$}]{ 
    \label{fig:subfig:a} %% label for first subfigure 
    \includegraphics[width=0.5\textwidth]{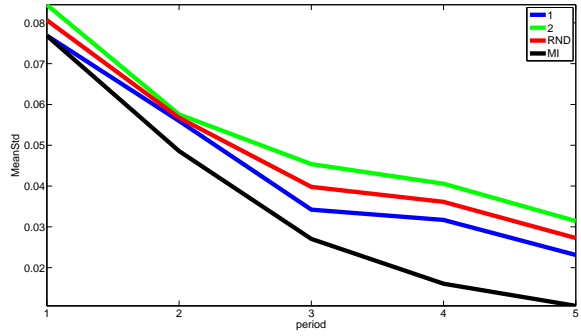}} 
		 \hspace{0.1in} 
	  \subfloat[\scriptsize{standard variance of $\theta_{2}$}]{ 
    \label{fig:subfig:b} %% label for first subfigure 
    \includegraphics[width=0.5\textwidth]{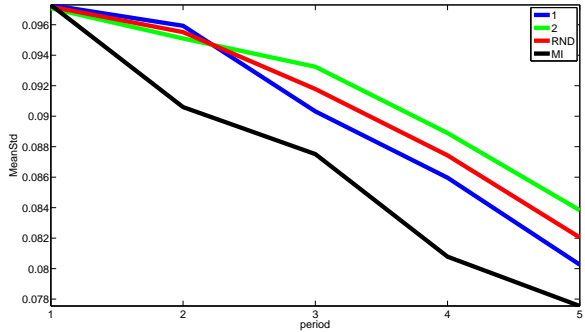}} 
		 \hspace{0.1in} 
	  \subfloat[\scriptsize{standard variance of $\theta_{3}$}]{ 
    \label{fig:subfig:c} %% label for first subfigure 
    \includegraphics[width=0.5\textwidth]{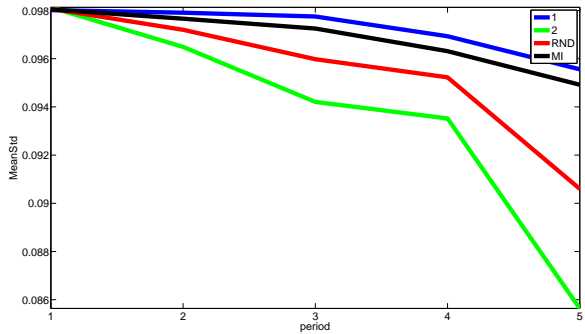}}  
		 \hspace{0.1in} 
  \subfloat[\scriptsize{joint entropy}]{ 
    \label{fig:subfig:d} %% label for second subfigure 
    \includegraphics[width=0.5\textwidth]{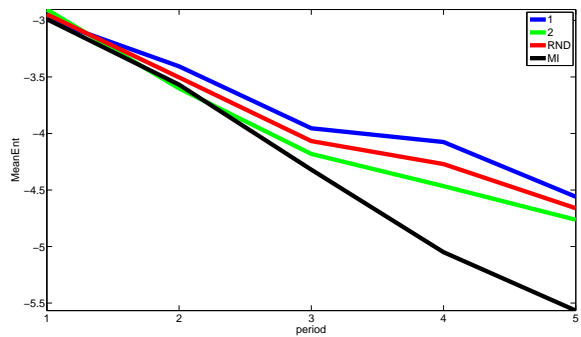}}
		\captionsetup{font={scriptsize}}
  \caption{The performance of maximum mutual information strategy and non-adaptive strategies are compared. Figure (a) (b) (c) show the standard variance of $\theta_{1}$,$\theta_{2}$ and $\theta_{3}$ after each experiment of different sequential designs. Figure (d) shows the joint entropy of three parameters which quantifies the total uncertainty of parameters.} 
  \label{fig:3parameters} %% label for entire figure 
\end{figure}

We use $2000$ ensemble members in the EnKF and to estimate the mutual information, and the performance measures are averaged over $100$ trial runs. The results are presented in Fig.\ref{fig:3parameters}. We can see that while the reduction in uncertainty of $\theta_{2}$ and $\theta_{3}$ are minor, maximum mutual information strategy gives the best estimate of $\theta_{1}$ and the smallest joint entropy of the three parameters as compared with the other three strategies. 

%We can see that strategy $1$(always choosing the first measurement time) and maximum mutual information strategy choose the same measurement time points and have the most reduction in uncertainty of parameters.

%----------------------------

\section{CONCLUSIONS}
\label{sec:conclusions}

In this paper, sequential experimental design is formulated in the Bayesian framework where the posterior distribution of model parameters obtained from one experiment is used as the prior information to design the next experiment. The design criterion is maximizing the expected information gain, which leads to the most reduction in parametric uncertainty. This design strategy is applied to two biological systems which are both represented by ordinary differential equations and it is showed that maximum mutual information strategy outperforms non-adaptive design strategies. However, several aspects needs improving and are planned as future work. First, in this paper, we use ensemble Kalman filter to perform Bayesian update, which is based on the assumption that the joint distribution between state and observable is Gaussian, which is not necessarily true on most occasions. Markov Chain Monte Carlo and sequential Monte Carlo methods have their weaknesses too. Thus we are looking for more accurate and general Bayesian update methods. Second, although mutual information is a powerful tool to measure the dependence between two random variables, estimating it is challenging and biased. Better ways to make a tradeoff between bias and variance are also needed. Third, on most occasions, we are unable to access the true model, so we have to consider how to deal with the model error.


%%%%%%%%%%%%%%%%%%%%%%%%%%%%%%%%%%%%%%%%%%%%%%%%%%%%%%%%%%%%%%%%%%%%%%%%%%%%%%%%

%%%%%%%%%%%%%%%%%%%%%%%%%%%%%%%%%%%%%%%%%%%%%%%%%%%%%%%%%%%%%%%%%%%%%%%%%%%%%%%%

%%%%%%%%%%%%%%%%%%%%%%%%%%%%%%%%%%%%%%%%%%%%%%%%%%%%%%%%%%%%%%%%%%%%%%%%%%%%%%%%
%\section*{APPENDIX}


%\section*{ACKNOWLEDGMENT}

%The preferred spelling of the word \D2acknowledgment\D3 in America is without an \D2e\D3 after the \D2g\D3. Avoid the stilted expression, \D2One of us (R. B. G.) thanks . . .\D3  Instead, try \D2R. B. G. thanks\D3. Put sponsor acknowledgments in the unnumbered footnote on the first page.

%%%%%%%%%%%%%%%%%%%%%%%%%%%%%%%%%%%%%%%%%%%%%%%%%%%%%%%%%%%%%%%%%%%%%%%%%%%%%%%%

%References are important to the reader; therefore, each citation must be complete and correct. If at all possible, references should be commonly available publications.

\bibliographystyle{IEEEtran}
\bibliography{ExpDesign}

\end{document}